%% file: main.tex
\documentclass[sigconf]{acmart}
\usepackage{pifont}
\AtBeginDocument{%
}
\copyrightyear{2026}
\acmYear{2026}
\setcopyright{cc}
\setcctype{by}
\acmConference[MM '26]{Proceedings of the 34th ACM International Conference on Multimedia}{November 10--14, 2026}{Rio de Janeiro, Brazil}
\acmBooktitle{Proceedings of the 34th ACM International Conference on Multimedia (MM '26), November 10--14, 2026, Rio de Janeiro, Brazil}
\acmDOI{10.1145/3767308.3836631}
\acmISBN{979-8-4007-2213-4/2026/11}

\begin{document}

\title{Beyond Speech: Dual-Domain SSL Fusion for Unified All-Type Audio Deepfake Detection}
\renewcommand{\shorttitle}{Dual-Domain SSL Fusion for Audio Deepfake Detection}

\author{Cunhang Fan}
\affiliation{%
  \department{State Key Laboratory of Opto-Electronic Information Acquisition and Protection Technology, School of Computer Science and Technology}
  \institution{Anhui University}
  \city{Hefei}
  \state{Anhui}
  \country{China}
}
\email{cunhang.fan@ahu.edu.cn}

\author{Junqin Cao}
\affiliation{%
  \department{State Key Laboratory of Opto-Electronic Information Acquisition and Protection Technology, School of Computer Science and Technology}
  \institution{Anhui University}
  \city{Hefei}
  \state{Anhui}
  \country{China}
}
\email{e24301282@stu.ahu.edu.cn}

\author{Tian Gao}
\affiliation{%
  \institution{Anhui Laboratory for Safe Artificial Intelligence in the Yangtze River Delta}
  \city{Hefei}
  \state{Anhui}
  \country{China}
}
\email{tiangao5@iflytek.com}

\author{Zhipeng Xie}
\affiliation{%
  \institution{Anhui Laboratory for Safe Artificial Intelligence in the Yangtze River Delta}
  \city{Hefei}
  \state{Anhui}
  \country{China}
}
\email{zpxie2@iflytek.com}

\author{Jun Xue}
\affiliation{%
  \department{Key Laboratory of Aerospace Information Security and Trusted Computing, Ministry of Education, School of Cyber Science and Engineering}
  \institution{Wuhan University}
  \city{Wuhan}
  \state{Hubei}
  \country{China}
}
\email{junxue@whu.edu.cn}

\author{Zhao Lv}
\affiliation{%
  \department{State Key Laboratory of Opto-Electronic Information Acquisition and Protection Technology, School of Computer Science and Technology}
  \institution{Anhui University}
  \city{Hefei}
  \state{Anhui}
  \country{China}
}
\email{kjlz@ahu.edu.cn}

\author{Xin Fang}
\authornote{Corresponding author.}
\affiliation{%
  \institution{University of Science and Technology of China}
  \city{Hefei}
  \state{Anhui}
  \country{China}
}
\email{xinfang@iflytek.com}

\renewcommand{\shortauthors}{Cunhang Fan et al.}

\begin{abstract}
Unified all-type audio deepfake detection aims to determine whether an input clip is real or fake when its audio type may be speech, environmental sound, singing voice, or music. Existing speech-centric or type-dependent solutions are insufficient for this setting because the test-time audio type is unknown, while the required output is still a single binary decision. To address these issues, this paper proposes a dual-domain SSL fusion method that maps heterogeneous audio into a shared binary authenticity space. EAT-large and wav2vec 2.0 XLS-R-300M are used as complementary SSL feature sources, providing broad acoustic and event-level representations as well as waveform-level, vocal, and speech-sensitive representations. Layer-wise weighted fusion integrates multi-level artifacts from different transformer depths, while token-level fusion forms a unified feature pool without enforcing frame-level alignment between the two SSL streams. The fused tokens are summarized by multi-head attentive statistics pooling and classified with a binary MLP head. With conservative speech refinement applied on top of this unified core detector, the submitted system achieves 95.58\% Macro-F1 on the AT-ADD Track 2 evaluation set and ranks second in the challenge.
\end{abstract}



\makeatletter
\let\savedTitleFootnoteSymbol\@fnsymbol
\renewcommand*{\@fnsymbol}[1]{%
  \ifnum#1=1\relax\protect\ding{41}\else\savedTitleFootnoteSymbol{#1}\fi}
\maketitle
\let\@fnsymbol\savedTitleFootnoteSymbol
\makeatother

\input{sections/01-introduction}
\input{sections/02-method}
\input{sections/03-experiments}
\input{sections/04-conclusion}

\begin{acks}
This work is supported by the National Natural Science Foundation of China (NSFC) (No.62571002, 62476004), Excellent Youth Foundation of Anhui Scientific Committee (No. 2408085Y034).
\end{acks}

\bibliographystyle{ACM-Reference-Format}
\bibliography{references}

\end{document}

%% file: sections/01-introduction.tex

\section{Introduction}

Recent advances in speech synthesis, general-audio generation, singing voice synthesis, and music generation have significantly improved the realism and accessibility of synthetic audio, raising new risks for multimedia security and digital trust. Audio deepfake detection therefore needs to move beyond speech spoofing and address heterogeneous audio content, including environmental sound, singing voice, and music. AT-ADD Track 2 formalizes this all-type setting by requiring a binary real/fake decision for each input clip whose audio type is unknown at inference time~\cite{atadd2026}. The task is thus not to recognize the audio category itself, but to build a detector whose binary decision remains reliable across different audio domains.

Speech-oriented benchmarks established controlled spoofing evaluation through ASVspoof~\cite{wu15e_interspeech,kinnunen17_interspeech,todisco19_interspeech,yamagishi21_asvspoof,wang24_asvspoof}, while the ADD series broadened evaluation to more diverse generation and recording conditions~\cite{add2022,yi2023add2023secondaudio}. Resulting detectors include spectrogram and raw-waveform systems~\cite{FAN2023102988,FAN2024106320}, graph attention networks~\cite{aasist}, and domain-general SSL approaches~\cite{xie2023singledomain,xie2023w2vasdg,xie2024asdg}. Related studies address real-time and social-media forensics, source tracing, and speech editing~\cite{10096837,10605999,10506099,xie2026fsw,xie2024refd,xue2026rtcfakespeechdeepfakedetection,xue2026unifyingspeecheditingdetection}. Speech SSL models such as wav2vec 2.0, XLS-R, and WavLM provide speaker, phonetic, and linguistic representations that are effective for speech spoofing~\cite{NEURIPS2020_92d1e1eb,babu22_interspeech,9814838,tak2022automaticspeakerverificationspoofing}.

All-type detection additionally requires non-speech representations. General-audio SSL models capture acoustic events and background scenes~\cite{yin25_interspeech,11464115,ijcai2024p421,beats,huang2026envtricascade}, while singing and music studies expose vocal-generation, song-level, and synthetic-music artifacts~\cite{svdd,xie2024fsd,xwsb,fakemusiccaps,xie2026wpt}. These lines are typically developed under domain-specific assumptions; AT-ADD instead requires one binary decision boundary across heterogeneous audio whose type is unknown at test time.

To address these issues, this paper proposes a dual-domain SSL fusion system for unified all-type audio deepfake detection. The core detector uses EAT-large and wav2vec 2.0 XLS-R-300M as complementary SSL feature sources. EAT-large contributes broad acoustic and event-level representations, whereas XLS-R-300M contributes waveform-level, vocal, and speech-sensitive representations. Instead of treating them as separate detectors, both streams are optimized toward the same shared binary authenticity space. Learnable layer-wise weighted fusion aggregates hidden states from multiple transformer layers, allowing the model to use low-level spectral artifacts, intermediate acoustic or event inconsistencies, and higher-level semantic cues~\cite{9688093,10096149}. The layer-fused token sequences are then concatenated along the temporal dimension to form a unified feature pool, which avoids forcing frame-level alignment between SSL streams. A SwiGLU projection module, multi-head attentive statistics pooling~\cite{india19_interspeech}, and an MLP binary classifier finally map the unified feature pool to real/fake logits. Every input follows this type-agnostic core detector. The final challenge submission retains the unified detectors for all samples and conditionally adds a speech-specific refiner only when two auxiliary cues agree that recognizable speech is present.

The main contributions of this work are summarized as follows:
\begin{itemize}
\item A dual-domain framework integrates complementary EAT-large and XLS-R representations into a shared binary authenticity space for unified all-type audio deepfake detection.
\item Layer-wise weighted fusion and token-level fusion build a unified feature pool for multi-head attentive statistics pooling without requiring frame-level alignment between the two SSL streams, while retaining each front-end's native token order and complementary temporal evidence during aggregation.
\item A conservative speech-refinement strategy is added on top of the unified core detector; the submitted system achieves 95.58\% Macro-F1 on the AT-ADD Track 2 evaluation set and ranks second in the challenge.
\end{itemize}

%% file: sections/02-method.tex

\section{Method}

\subsection{Architecture}

\begin{figure*}[t]
\centering
\includegraphics[width=0.98\textwidth,trim=0 12bp 0 0,clip]{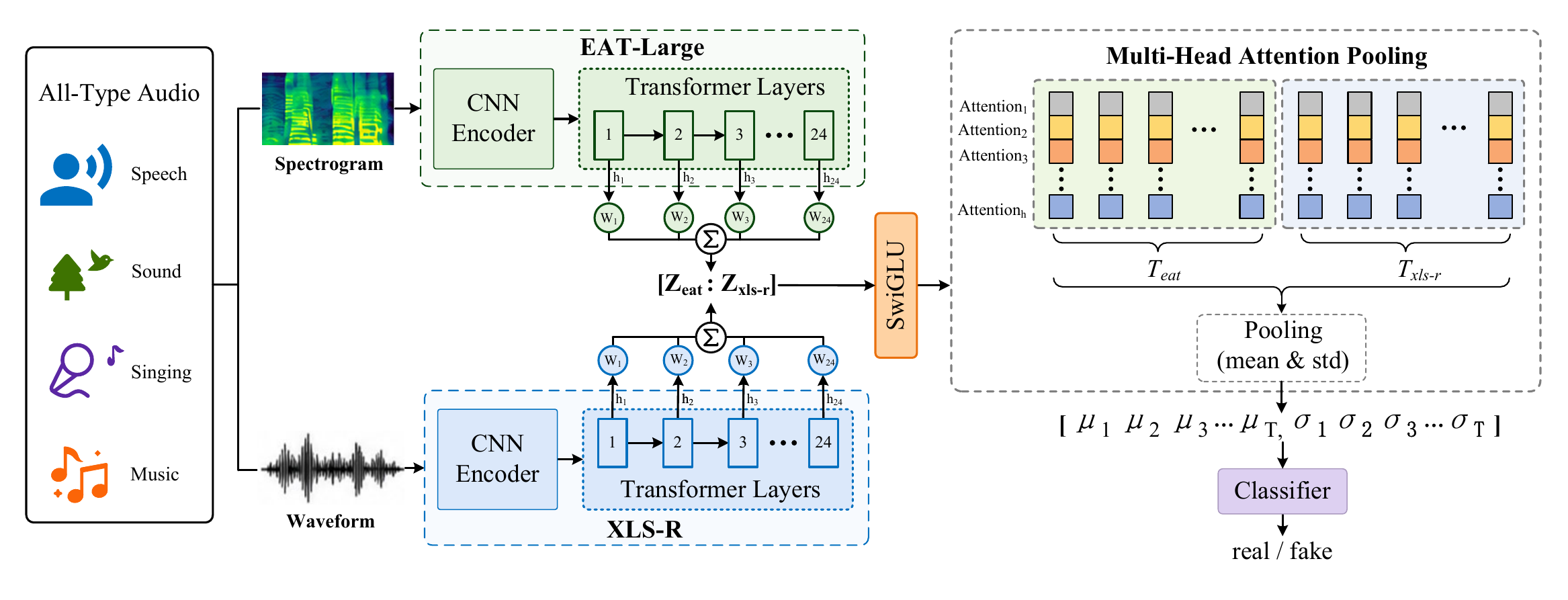}
\Description{Dual-domain audio deepfake detector with EAT-large and XLS-R streams, layer-wise fusion, SwiGLU projection, multi-head attentive statistics pooling, and an MLP real/fake classifier.}
\caption{Architecture of the proposed dual-domain all-type audio deepfake detector. EAT-large and XLS-R extract layer-wise fused representations from spectrogram and waveform inputs, respectively; the concatenated token sequence is processed by SwiGLU projection, multi-head attentive statistics pooling, and an MLP binary classifier.}\label{fig:architecture}
\end{figure*}

The core detector follows a single type-agnostic path for speech, environmental sound, singing voice, and music. As shown in Figure~\ref{fig:architecture}, EAT-large and XLS-R-300M front-ends, learnable layer-wise fusion, token-level fusion, multi-head attentive statistics pooling, and an MLP binary classifier map heterogeneous inputs into a shared real/fake decision space.

\subsection{Dual-Domain SSL Feature Fusion}

All-type audio deepfake detection covers speech, environmental sound, singing voice, and music, whose artifacts may appear in linguistic, vocal, acoustic-event, timbral, or temporal structures. A speech-oriented SSL model emphasizes vocal and linguistic structure, whereas a general-audio SSL model provides complementary acoustic-event and non-linguistic cues. We therefore use EAT-large and XLS-R-300M as complementary front-ends for one unified detector. EAT-large takes time-frequency features and provides broad acoustic and event-level representations, whereas XLS-R-300M takes the raw waveform and provides waveform-level, vocal, and speech-sensitive representations. Both streams are optimized toward the same shared binary authenticity space rather than separate type-specific decision paths.

Transformer SSL layers encode cues at different abstraction levels. Low layers preserve local acoustic and spectral details, intermediate layers encode phonetic or event-level patterns, and high layers capture more abstract semantics. Since deepfake artifacts may occur at any of these levels, using only the final layer can discard useful cues. We thus apply learnable layer-wise weighted fusion to both EAT-large and XLS-R-300M.

Let $H_l^m$ denote the hidden representation extracted from the $l$-th transformer layer of model $m$, where $m \in {\mathrm{EAT}, \mathrm{XLSR}}$. For each model, a learnable scalar weight $w_l^m$ is assigned to each layer and normalized by a softmax function:
\begin{equation}
\alpha_l^m =
\frac{\exp(w_l^m)}
{\sum_{i=1}^{L_m}\exp(w_i^m)} ,
\end{equation}
where $L_m$ denotes the number of transformer layers. The layer-fused representation is computed as:
\begin{equation}
H^m =
\sum_{l=1}^{L_m}
\alpha_l^m H_l^m.
\end{equation}

In our implementation, both EAT-large and XLS-R-300M contain 24 transformer layers. The layer weights are learned independently for the two streams, allowing different abstraction levels to be selected under the same binary authenticity objective. This design is useful because the two SSL models do not necessarily expose the most discriminative authenticity cues at the same transformer depth.

After layer-wise fusion, let $E \in \mathbb{R}^{T_E \times D}$ and $X \in \mathbb{R}^{T_X \times D}$ denote the EAT and XLS-R token sequences, where $T_E$ and $T_X$ are token lengths and $D$ is their shared feature dimension. Instead of concatenating the two sequences along the feature dimension, we concatenate them along the temporal or token dimension:
\begin{equation}
Z = [E; X] \in \mathbb{R}^{(T_E+T_X) \times D}.
\end{equation}

Both pretrained encoders expose the same hidden dimension $D$, so their layer-fused tokens are concatenated directly, without an additional stream-specific projection, cross-stream normalization, temporal resampling, or frontend-identity embedding. The shared SwiGLU layer is the first learned transformation of the combined pool.

Token-level concatenation avoids assuming frame-level alignment between SSL streams with different inputs and pretraining objectives. Feature-dimensional concatenation would require one-to-one correspondence between EAT and XLS-R tokens, which is not guaranteed because their front-ends, token rates, and learned semantics differ. By concatenating along the token dimension, the model preserves each stream's internal structure while forming a unified pool. Attention is normalized over the complete sequence, allowing pooling to learn token-dependent relevance from content without a test-time audio type label.

This fusion strategy also keeps the core binary classifier independent of audio-type routing. All EAT and XLS-R tokens are exposed to the same pooling and classification modules, so the core detector does not need to decide whether a clip should follow a speech, music, sound, or singing path. Instead, the learned representation is organized around real/fake discrimination. Audio type can still be useful for auxiliary supervision and analysis; the conservative refinement used by the final challenge submission is described separately in Section~\ref{sec:ensemble}.

A lightweight projection based on SwiGLU~\cite{shazeer2020gluvariantsimprovetransformer} processes the concatenated sequence before attention aggregation. This gated nonlinear layer enhances useful responses and suppresses less informative activations.

The projected token sequence contains both EAT-derived and XLS-R-derived tokens. To summarize this unified feature pool, we use multi-head attentive statistics pooling~\cite{india19_interspeech}. Let $\tilde{Z}\in \mathbb{R}^{D\times T}$ denote the projected sequence, where $T=T_E+T_X$. The feature dimension is divided into $H$ heads. Each token is written as $\tilde{z}_t=[\tilde{z}_{1,t};\ldots;\tilde{z}_{H,t}]$, where $\tilde{z}_{h,t}\in\mathbb{R}^{D/H}$.

For each head, an attention component computes a scalar score from the corresponding feature split and normalizes the scores over time:
\begin{equation}
e_{h,t}=g_h(\tilde{z}_{h,t}), \quad
a_{h,t}=
\frac{\exp(e_{h,t})}
{\sum_{\tau=1}^{T}\exp(e_{h,\tau})},
\end{equation}
where $g_h(\cdot)$ is implemented by a grouped one-dimensional affine attention component. The weights are therefore temporal importance scores within each feature head, rather than scaled dot-product self-attention weights.

The attention scores select authenticity-relevant cues from the unified feature pool. Different heads can emphasize different SSL feature sources, temporal regions, or artifact patterns, while all heads still contribute to the same type-agnostic binary real/fake decision.

For the $h$-th attention head, the weighted mean representation is computed as:
\begin{equation}
\mu_h =
\sum_{t=1}^{T}
a_{h,t} \tilde{z}_{h,t} ,
\end{equation}

In addition to the weighted mean, we also compute the weighted standard deviation:
\begin{equation}
\sigma_h =
\sqrt{\sum_{t=1}^{T} a_{h,t}
{(\tilde{z}_{h,t}-\mu_h)}^2},
\end{equation}
where the square is applied element-wise. In implementation, this is computed equivalently as the attention-weighted second moment minus the squared mean.

The final utterance-level representation is obtained by concatenating the mean and standard deviation statistics from all attention heads:
\begin{equation}
r =
[\mu_1;\ldots;\mu_H;\sigma_1;\ldots;\sigma_H],
\end{equation}
where $H$ is the number of attention heads.

This statistical pooling step is useful because fake artifacts may be sparse, local, or distributed unevenly across time. The weighted mean captures the dominant discriminative cues, while the weighted standard deviation reflects the variation and instability of the selected tokens. Therefore, the model can represent not only which cues are present, but also how consistently they appear across the fused EAT--XLS-R token sequence.

The pooled representation is finally passed to an MLP classifier to produce binary real/fake logits $\mathbf{z} \in \mathbb{R}^{2}$.

%% file: sections/03-experiments.tex

\section{Experiments}

\subsection{Experimental Setup}

AT-ADD Track 2 evaluates binary real/fake classification across speech, environmental sound, singing voice, and music. Audio type is unavailable during inference and is not a prediction target. The official metric first computes binary Macro-F1 for each audio type and then averages the four type-wise scores. These scores support evaluation and diagnosis, whereas the submitted system produces only one type-agnostic authenticity decision per clip.

We use only the officially released training and development sets, which contain 146,781 and 91,069 samples, respectively. After merging them into a pool of 237,850 samples, we construct five stratified train/validation configurations with the random seed fixed to 42. For configuration $k$, one fold containing 47,570 samples is used as the internal validation set $\mathrm{dev}_k$, and the remaining four folds containing 190,280 samples form $\mathrm{train}_k$. Stratification preserves the joint distribution of audio type, real/fake label, and generator identity; generator identity is a stratification attribute rather than a grouping constraint, so the same generator may appear in both $\mathrm{train}_k$ and $\mathrm{dev}_k$. Each internal validation set is used only for early stopping and checkpoint selection of its corresponding model. The official progress and evaluation sets are outside these five configurations and are not used for gradient updates, early stopping, checkpoint selection, pseudo-labeling, self-training, or threshold tuning.

The five internal validation folds form a disjoint partition of the merged pool, and $\mathrm{train}_k$ is the complement of $\mathrm{dev}_k$. Thus, every labeled sample is used for internal validation once and internal training four times, with no train--validation sample overlap within a configuration. Only the partition changes; the architecture, optimization recipe, and fixed training seed remain the same.

All models are implemented in PyTorch. During training, a random 4-second crop is sampled from each audio clip. During inference, we use five-crop test-time averaging with 4-second segments. We train each model for up to 50 epochs with an early-stopping patience of 5. AdamW is used as the optimizer~\cite{loshchilov2017decoupled}. The SSL front-ends are optimized with a learning rate of $1\times10^{-6}$, while the newly initialized classification heads are optimized with a learning rate of $1\times10^{-5}$.

For the all-type EAT--XLS-R detectors, we apply RawBoost Algorithm 5 during training~\cite{9746213}. For the speech-specific auxiliary refiner, we use stronger speech-oriented augmentation, including M4A re-encoding, MUSAN-based additive noise, and RIRS-based reverberation. The M4A bitrates are randomly sampled from 12, 16, 24, 32, and 48 kbps. MUSAN augmentation uses noise, music, and speech sources with signal-to-noise ratios ranging from 5 dB to 15 dB.

\subsection{System Configuration and Ensemble Strategy}\label{sec:ensemble}

Table~\ref{tab:system_config} summarizes all models considered in our system. These models can be divided into three groups: unified all-type detectors, diagnostic specialization baselines, and auxiliary cue models. S1 is an EAT-large all-type baseline. S2, S3, and S4 are all-type EAT--XLS-R detectors with concatenation, additive fusion, and cross-attention fusion, respectively. S5 extends S2 by adding an auxiliary audio type classification head, so it produces both binary real/fake logits and type predictions. S6 and S7 are diagnostic binary baselines trained on coarse subsets: S6 uses human vocal audio, including speech and singing voice, while S7 uses non-vocal audio, including environmental sound and music. S8 is an EAT--XLS-R auxiliary type classifier based on concatenation fusion. S9 is a frozen Whisper-large-v3 model used only to test whether an input yields a nonempty ASR transcription; it is not fine-tuned and produces no real/fake score. S10 is a speech-specific auxiliary refiner trained exclusively on speech samples.

\begin{table}[t]
\caption{Configuration of all models considered in our system.}\label{tab:system_config}
\centering
\resizebox{\columnwidth}{!}{
\begin{tabular}{lll}
\toprule
ID & System & Train \\
\midrule
S1 & EAT-large + binary head & All \\
S2 & EAT-large + XLS-R-300M (Cat) + binary head & All \\
S3 & EAT-large + XLS-R-300M (Add) + binary head & All \\
S4 & EAT-large + XLS-R-300M (Cross-attn.) + binary head & All \\
S5 & S2 + type head & All \\
S6 & XLS-R-300M + binary head & Speech, Singing \\
S7 & EAT-large + binary head & Sound, Music \\
S8 & EAT-large + XLS-R-300M (Cat) + type head & All \\
S9 & Whisper-large-v3 & None \\
S10 & XLS-R-300M + binary head & Speech \\
\bottomrule
\end{tabular}
}
\end{table}

We include S6+S7+S8 as a diagnostic hard type-selection baseline. In this setting, S8 predicts the audio type, and the binary prediction is taken from S6 for speech or singing voice and from S7 for environmental sound or music. This baseline is used only to test whether explicit type-based selection helps; it is not the main modeling strategy because its binary decision depends on the correctness of an intermediate type prediction.

The final submitted system keeps the unified all-type detectors as the main path for every input. Let $q_5=1$ when S5 predicts the speech type and $q_9=1$ when the frozen S9 model returns a nonempty transcription. S9 neither predicts authenticity nor contributes a logit. The speech-refinement gate is $q=q_5q_9$, and the final logit is
\begin{equation}
\mathbf{z}_{\mathrm{final}}=
\begin{cases}
(\mathbf{z}_2+\mathbf{z}_5+\mathbf{z}_{10})/3, & q=1,\\
(\mathbf{z}_2+\mathbf{z}_5)/2, & q=0.
\end{cases}
\end{equation}
A fixed threshold of 0.5 is applied to the real-class probability obtained from $\mathbf{z}_{\mathrm{final}}$. Thus, S2 and S5 remain active for every sample, while S10 is conditionally added as a conservative speech-specific refiner. This refinement is type-aware, but it does not replace the unified all-type detection path.

\subsection{Results}

We first report the performance of S1 under the five train/validation configurations. For each configuration, training, early stopping, and checkpoint selection are completed using only $\mathrm{train}_k$ and $\mathrm{dev}_k$. The resulting five frozen S1 models are then submitted independently to the official progress leaderboard. Table~\ref{tab:s1_progress_folds} reports the returned per-type Macro-F1 scores over speech, sound, singing, and music, together with their final average. The differences across rows reflect the five data partitions rather than repeated random-seed trials on a fixed split.

\begin{table}[t]
\caption{Macro-F1 (\%) of S1 across five split configurations on the progress set.}\label{tab:s1_progress_folds}
\centering
\resizebox{\columnwidth}{!}{
\begin{tabular}{lccccc}
\toprule
System & Speech & Sound & Singing & Music & Macro-F1 \\
\midrule
S1 config. 1 & 79.75 & \textbf{99.07} & 94.34 & \textbf{98.67} & \textbf{92.96} \\
S1 config. 2 & \textbf{80.66} & 98.50 & \textbf{95.18} & 96.83 & 92.79 \\
S1 config. 3 & 79.42 & 97.90 & 93.22 & 93.28 & 90.95 \\
S1 config. 4 & 77.47 & 98.54 & 94.22 & 96.66 & 91.72 \\
S1 config. 5 & 76.42 & 96.32 & 92.87 & 96.03 & 90.41 \\
\bottomrule
\end{tabular}
}
\end{table}

Based only on the aggregate scores returned by the official progress leaderboard, configuration 1 is fixed as the common training and validation split for subsequent system comparisons. No progress-set audio, labels, or sample-level predictions are accessed, and the returned scores are not used for parameter updates, within-fold checkpoint selection, or threshold tuning. Configuration 1 is retained because its S1 model provides the highest averaged Macro-F1 and a strong balance across the four diagnostic audio types.

Thus, the progress leaderboard is used once to choose the data partition carried forward, not to choose among epochs or checkpoints. After configuration 1 is fixed, all later variants use the same $\mathrm{train}_1$/$\mathrm{dev}_1$ roles, avoiding a split confound in their progress-set comparison.

We then summarize the progress-set performance of all experimental systems in Table~\ref{tab:progress_all_systems}. This comparison covers the EAT-large baseline, different EAT--XLS-R fusion strategies, the auxiliary type-head detector, and the hard type-selection baseline. It allows us to compare unified all-type detection, auxiliary type supervision, and explicit type-based selection under the same progress-set evaluation protocol. In this table, S1 is trained with the original official train/dev split, while all other systems use split configuration 1. DA denotes RawBoost Algorithm 5; unless otherwise specified, all systems use this augmentation. S2* denotes an S2 variant that applies additional speech-oriented augmentation for speech samples.

\begin{table}[t]
\caption{Macro-F1 (\%) of all experimental systems on the progress set.}\label{tab:progress_all_systems}
\centering
\resizebox{\columnwidth}{!}{
\begin{tabular}{lccccc}
\toprule
System & Speech & Sound & Singing & Music & Macro-F1 \\
\midrule
S1 w/o DA & 74.79 & 88.25 & 95.98 & 65.37 & 81.10 \\
S1 & 79.47 & 98.93 & 92.40 & \textbf{98.45} & 92.31 \\
S2 & \textbf{84.38} & 98.18 & 97.39 & 98.25 & \textbf{94.55} \\
S2* & 80.52 & 97.12 & 97.01 & 97.59 & 93.06 \\
S3 & 81.91 & 99.57 & 97.30 & 98.38 & 94.29 \\
S4 & 82.26 & \textbf{99.61} & 96.87 & 97.93 & 94.17 \\
S5 & 84.00 & 97.85 & \textbf{98.15} & 97.82 & 94.45 \\
S6+S7+S8 & 78.66 & 97.77 & 93.98 & 95.43 & 91.46 \\
\bottomrule
\end{tabular}
}
\end{table}

Finally, Table~\ref{tab:eval_final_fusion} reports evaluation-set performance. S2 and S5 both use split configuration 1. In the intermediate S2+S9+S10 variant, a nonempty S9 transcription activates the average of S2 and S10; otherwise, S2 is used alone. The final row uses the stricter S5--S9 intersection gate defined above and averages S2, S5, and S10 only for the selected speech-refinement subset. Relative to S2, the final fusion improves speech, sound, singing, and music by 1.31, 1.09, 1.33, and 0.87 points, respectively, yielding a 1.14-point Macro-F1 gain. The improvement is therefore distributed across all four diagnostic categories rather than concentrated in speech.

\begin{table}[t]
\caption{Macro-F1 (\%) of the baseline, S2, and fusion systems on the evaluation set.}\label{tab:eval_final_fusion}
\centering
\resizebox{\columnwidth}{!}{
\begin{tabular}{lccccc}
\toprule
System & Speech & Sound & Singing & Music & Macro-F1 \\
\midrule
Baseline & 79.47 & 79.5 & 66.82 & 96.3 & 75.28 \\
S2 & 83.92 & 97.81 & 98.14 & 97.87 & 94.44 \\
S2+S9+S10 & \textbf{85.28} & 97.81 & 98.15 & 97.87 & 94.77 \\
S2+S5+S9+S10 & 85.23 & \textbf{98.90} & \textbf{99.47} & \textbf{98.74} & \textbf{95.58} \\
\bottomrule
\end{tabular}
}
\end{table}

\subsection{Analysis}

The type-wise scores are diagnostic axes rather than separate targets: each input receives one binary real/fake decision. The key question is therefore whether a shared authenticity space remains stable across heterogeneous audio, rather than whether each type requires an independent decision path.

\begin{figure}[t]
\centering
\includegraphics[width=0.93\linewidth]{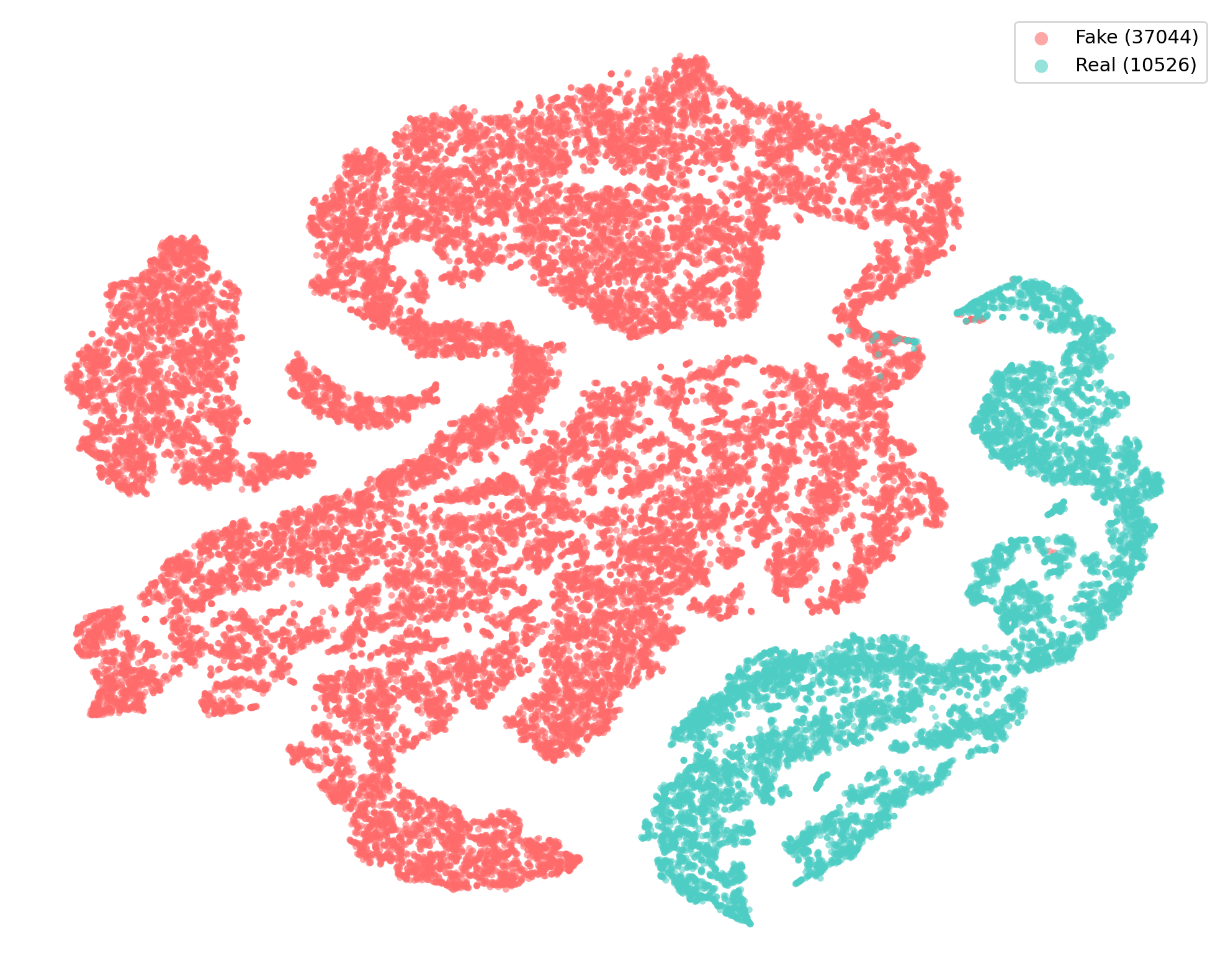}
\vspace{0.4ex}
\includegraphics[width=0.93\linewidth]{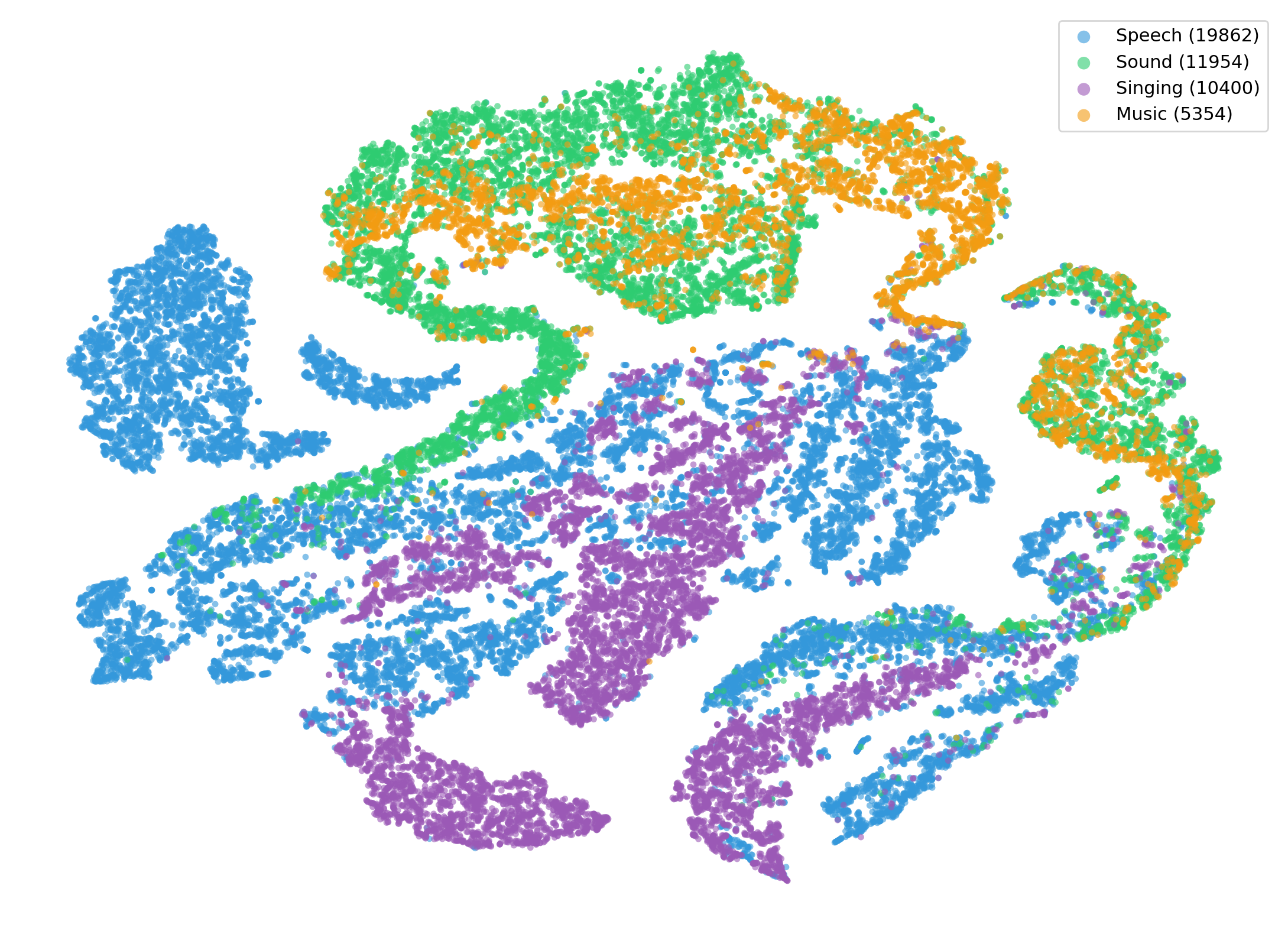}
\Description{t-SNE visualizations of binary-classification embeddings on the first internal validation set, colored by binary authenticity labels and audio types.}
\caption{t-SNE visualizations of binary-classification embeddings on $\mathrm{dev}_1$, colored by real/fake label and audio type.}\label{fig:tsne_visualizations}
\end{figure}

The t-SNE visualization in Figure~\ref{fig:tsne_visualizations} is computed on $\mathrm{dev}_1$ and is used only to inspect the learned binary-classification embedding. Coloring the same embedding by real/fake label shows a clear separation between authenticity labels, while coloring it by audio type reveals heterogeneous but jointly embedded audio domains. This supports the intended behavior of the unified core detector: audio type explains part of the data distribution, but the core representation is still optimized for a shared binary decision.

The progress-set comparison supports the effectiveness of unified dual-domain SSL features. The EAT--XLS-R systems improve over the EAT-only baseline, and the concatenation, additive fusion, and cross-attention variants all remain competitive. This trend is consistent with combining broad acoustic and event-level representations with waveform-level and speech-sensitive representations in a shared real/fake space. S5 remains close to S2 in standalone performance (94.45\% versus 94.55\% Macro-F1); its auxiliary type head therefore does not improve the standalone score, but it provides the type cue and complementary binary logits used by the final fusion.

The available XLS-R-only systems use speech-related subsets rather than all four audio types and therefore do not isolate EAT-large's per-type marginal contribution. Likewise, Table~\ref{tab:s1_progress_folds} measures partition sensitivity rather than random-seed variance because configuration 1 was selected by its progress score; the evaluation set remains the final blind comparison.

Data augmentation has different effects depending on whether it supports the all-type objective. RawBoost raises S1 Macro-F1 from 81.10\% to 92.31\%, mainly through large gains on sound and music, because it improves robustness to channel, codec, and acoustic perturbations and reduces overfitting to superficial spectral conditions. The singing decrease shows that the benefit is not uniform across all types, but the macro gain indicates that broad augmentation helps the shared binary decision space. By contrast, S2* drops from 94.55\% to 93.06\% after adding speech-oriented augmentation. With EAT--XLS-R fusion and RawBoost already providing robust features, augmenting only speech can disturb cross-type balance, so speech cues are better used for conservative refinement than for reshaping all-type training.

\begin{figure}[t]
\centering
\includegraphics[width=0.86\linewidth]{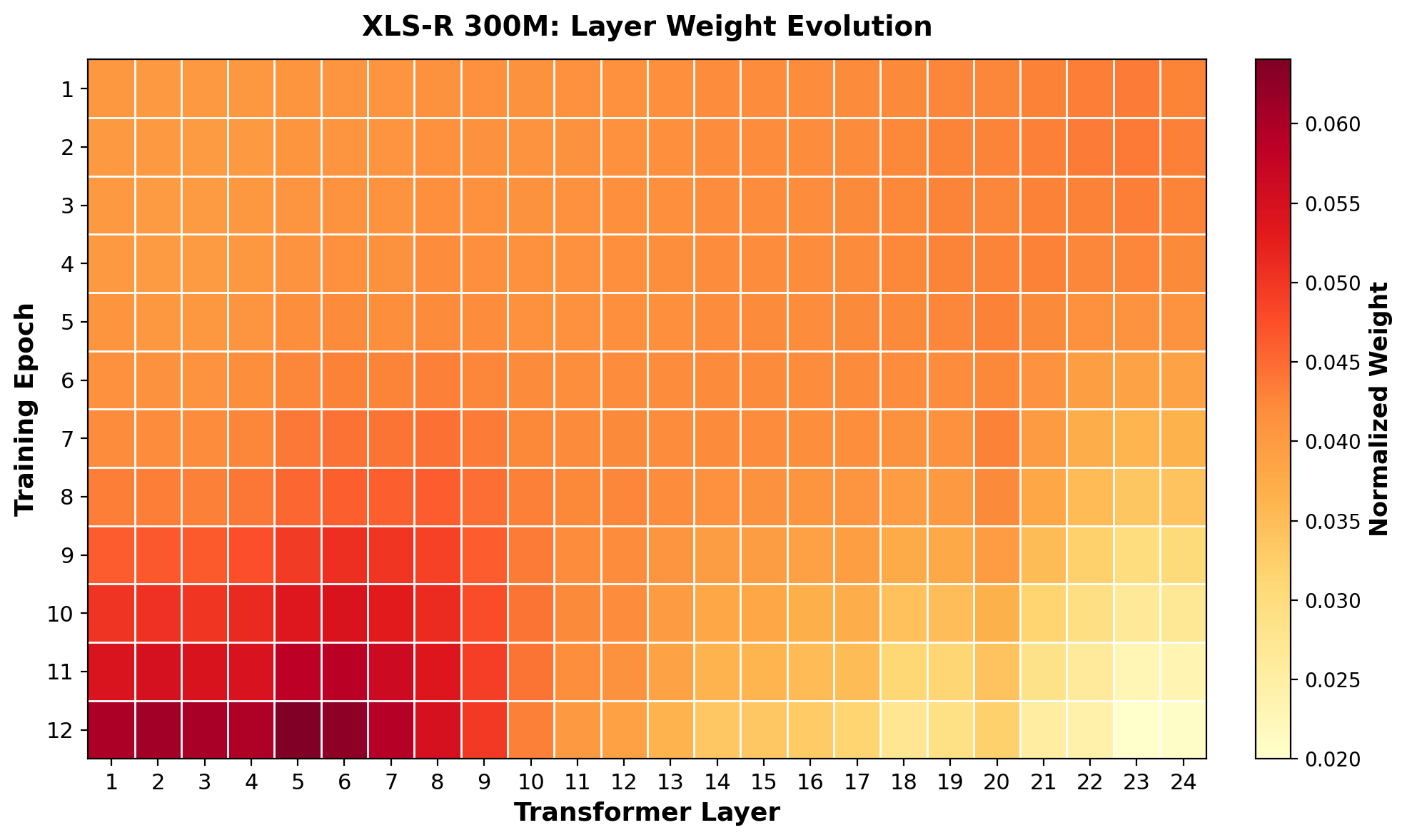}
\vspace{0.3ex}
\includegraphics[width=0.86\linewidth]{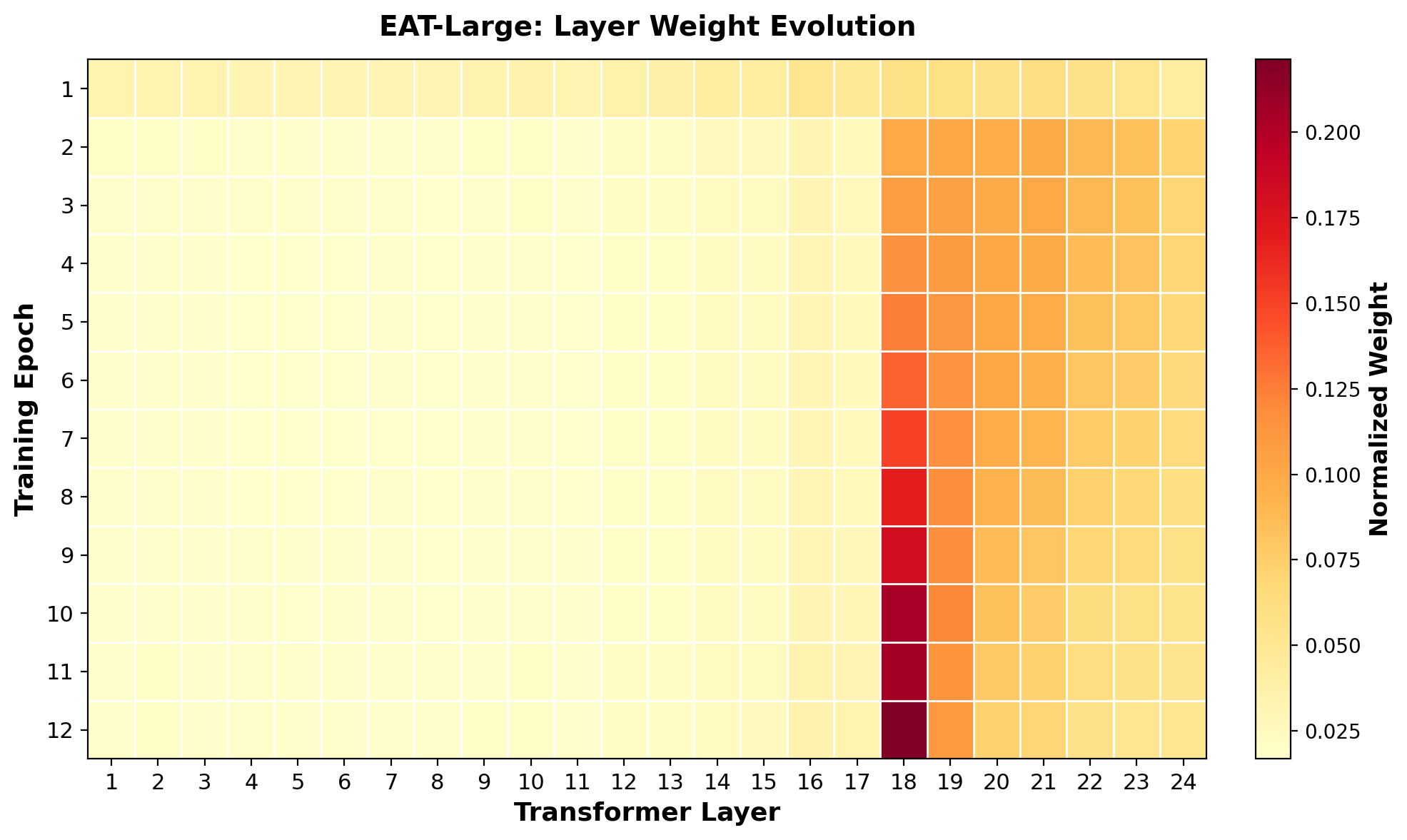}
\Description{Layer-weight heatmaps for XLS-R-300M and EAT-large over training epochs.}
\caption{Normalized layer weights during training for XLS-R-300M and EAT-large.}\label{fig:layer_weight_heatmaps}
\end{figure}

Figure~\ref{fig:layer_weight_heatmaps} illustrates that XLS-R-300M emphasizes lower and middle layers in later epochs, whereas EAT-large concentrates on a high-level layer. The independently learned patterns are consistent with the two streams contributing different abstraction levels under the shared binary objective.

S6+S7+S8's lower score (91.46\%) shows that hard routing propagates type-prediction errors, supporting a unified path with only conservative speech refinement.

%% file: sections/04-conclusion.tex

\section{Conclusion}
This paper presented a unified AT-ADD Track 2 detector. Layer-wise weighting and token-level fusion combine EAT-large and XLS-R-300M without frame alignment; shared SwiGLU, attentive statistics pooling, and a binary head retain one all-type path. The final submission conditionally adds a speech refiner using joint auxiliary-type and frozen-ASR cues, while S2 and S5 remain active for every sample.

The system ranks second with 95.58\% Macro-F1, 20.30 points above the baseline. Relative to S2, the final fusion gains 1.31, 1.09, 1.33, and 0.87 points on speech, sound, singing, and music. The results support dual-domain fusion with conservative refinement across heterogeneous audio. The embedding and layer-weight analyses further show complementary SSL cues within the shared authenticity space. Together, these results demonstrate that a shared detector can exploit type-sensitive cues without introducing type-dependent inference routes.

%% file: main.bbl

\begin{thebibliography}{41}


\ifx \showCODEN    \undefined \def \showCODEN     #1{\unskip}     \fi
\ifx \showISBNx    \undefined \def \showISBNx     #1{\unskip}     \fi
\ifx \showISBNxiii \undefined \def \showISBNxiii  #1{\unskip}     \fi
\ifx \showISSN     \undefined \def \showISSN      #1{\unskip}     \fi
\ifx \showLCCN     \undefined \def \showLCCN      #1{\unskip}     \fi
\ifx \shownote     \undefined \def \shownote      #1{#1}          \fi
\ifx \showarticletitle \undefined \def \showarticletitle #1{#1}   \fi
\ifx \showURL      \undefined \def \showURL       {\relax}        \fi
\providecommand\bibfield[2]{#2}
\providecommand\bibinfo[2]{#2}
\providecommand\natexlab[1]{#1}
\providecommand\showeprint[2][]{arXiv:#2}

\bibitem[Babu et~al\mbox{.}(2022)]%
        {babu22_interspeech}
\bibfield{author}{\bibinfo{person}{Arun Babu}, \bibinfo{person}{Changhan Wang},
  \bibinfo{person}{Andros Tjandra}, \bibinfo{person}{Kushal Lakhotia},
  \bibinfo{person}{Qiantong Xu}, \bibinfo{person}{Naman Goyal},
  \bibinfo{person}{Kritika Singh}, \bibinfo{person}{Patrick {von Platen}},
  \bibinfo{person}{Yatharth Saraf}, \bibinfo{person}{Juan Pino},
  \bibinfo{person}{Alexei Baevski}, \bibinfo{person}{Alexis Conneau}, {and}
  \bibinfo{person}{Michael Auli}.} \bibinfo{year}{2022}\natexlab{}.
\newblock \showarticletitle{{XLS-R}: Self-Supervised Cross-Lingual Speech
  Representation Learning at Scale}. In \bibinfo{booktitle}{\emph{Interspeech
  2022}}. \bibinfo{pages}{2278--2282}.
\newblock
\href{https://doi.org/10.21437/Interspeech.2022-143}{doi:\nolinkurl{10.21437/Interspeech.2022-143}}


\bibitem[Baevski et~al\mbox{.}(2020)]%
        {NEURIPS2020_92d1e1eb}
\bibfield{author}{\bibinfo{person}{Alexei Baevski}, \bibinfo{person}{Yuhao
  Zhou}, \bibinfo{person}{Abdelrahman Mohamed}, {and} \bibinfo{person}{Michael
  Auli}.} \bibinfo{year}{2020}\natexlab{}.
\newblock \showarticletitle{wav2vec 2.0: A Framework for Self-Supervised
  Learning of Speech Representations}. In \bibinfo{booktitle}{\emph{Advances in
  Neural Information Processing Systems}}, Vol.~\bibinfo{volume}{33}.
  \bibinfo{publisher}{Curran Associates, Inc.}, \bibinfo{pages}{12449--12460}.
\newblock


\bibitem[Cao et~al\mbox{.}(2026)]%
        {11464115}
\bibfield{author}{\bibinfo{person}{Junqin Cao}, \bibinfo{person}{Cunhang Fan},
  \bibinfo{person}{Jun Xue}, \bibinfo{person}{Yuankun Xie},
  \bibinfo{person}{Ruibo Fu}, \bibinfo{person}{Zhengqi Wen},
  \bibinfo{person}{Jiangyan Yi}, \bibinfo{person}{Yanzhen Ren},
  \bibinfo{person}{Zhao Lv}, {and} \bibinfo{person}{Jianhua Tao}.}
  \bibinfo{year}{2026}\natexlab{}.
\newblock \showarticletitle{Efficient Audio Transformer and {AASIST} for
  Environment Sound Deepfake Detection in the {ESDD} 2026 Challenge}. In
  \bibinfo{booktitle}{\emph{Proceedings of ICASSP}}.
  \bibinfo{pages}{21781--21783}.
\newblock
\href{https://doi.org/10.1109/ICASSP55912.2026.11464115}{doi:\nolinkurl{10.1109/ICASSP55912.2026.11464115}}


\bibitem[Chen et~al\mbox{.}(2022)]%
        {9814838}
\bibfield{author}{\bibinfo{person}{Sanyuan Chen}, \bibinfo{person}{Chengyi
  Wang}, \bibinfo{person}{Zhengyang Chen}, \bibinfo{person}{Yu Wu},
  \bibinfo{person}{Shujie Liu}, \bibinfo{person}{Zhuo Chen},
  \bibinfo{person}{Jinyu Li}, \bibinfo{person}{Naoyuki Kanda},
  \bibinfo{person}{Takuya Yoshioka}, \bibinfo{person}{Xiong Xiao},
  \bibinfo{person}{Jian Wu}, \bibinfo{person}{Long Zhou}, \bibinfo{person}{Shuo
  Ren}, \bibinfo{person}{Yanmin Qian}, \bibinfo{person}{Yao Qian},
  \bibinfo{person}{Michael Zeng}, \bibinfo{person}{Xiangzhan Yu}, {and}
  \bibinfo{person}{Furu Wei}.} \bibinfo{year}{2022}\natexlab{}.
\newblock \showarticletitle{{WavLM}: Large-Scale Self-Supervised Pre-Training
  for Full Stack Speech Processing}.
\newblock \bibinfo{journal}{\emph{IEEE Journal of Selected Topics in Signal
  Processing}} \bibinfo{volume}{16}, \bibinfo{number}{6}
  (\bibinfo{year}{2022}), \bibinfo{pages}{1505--1518}.
\newblock
\href{https://doi.org/10.1109/JSTSP.2022.3188113}{doi:\nolinkurl{10.1109/JSTSP.2022.3188113}}


\bibitem[Chen et~al\mbox{.}(2023)]%
        {beats}
\bibfield{author}{\bibinfo{person}{Sanyuan Chen}, \bibinfo{person}{Yu Wu},
  \bibinfo{person}{Chengyi Wang}, \bibinfo{person}{Shujie Liu},
  \bibinfo{person}{Daniel Tompkins}, \bibinfo{person}{Zhuo Chen},
  \bibinfo{person}{Wanxiang Che}, \bibinfo{person}{Xiangzhan Yu}, {and}
  \bibinfo{person}{Furu Wei}.} \bibinfo{year}{2023}\natexlab{}.
\newblock \showarticletitle{{BEATs}: Audio Pre-Training with Acoustic
  Tokenizers}. In \bibinfo{booktitle}{\emph{Proceedings of the 40th
  International Conference on Machine Learning}}. \bibinfo{publisher}{PMLR},
  \bibinfo{address}{Honolulu, HI, USA}, \bibinfo{pages}{5178--5193}.
\newblock


\bibitem[Chen et~al\mbox{.}(2024)]%
        {ijcai2024p421}
\bibfield{author}{\bibinfo{person}{Wenxi Chen}, \bibinfo{person}{Yuzhe Liang},
  \bibinfo{person}{Ziyang Ma}, \bibinfo{person}{Zhisheng Zheng}, {and}
  \bibinfo{person}{Xie Chen}.} \bibinfo{year}{2024}\natexlab{}.
\newblock \showarticletitle{{EAT}: Self-Supervised Pre-Training with Efficient
  Audio Transformer}. In \bibinfo{booktitle}{\emph{Proceedings of the
  Thirty-Third International Joint Conference on Artificial Intelligence}}.
  \bibinfo{pages}{3807--3815}.
\newblock
\href{https://doi.org/10.24963/ijcai.2024/421}{doi:\nolinkurl{10.24963/ijcai.2024/421}}


\bibitem[Comanducci et~al\mbox{.}(2024)]%
        {fakemusiccaps}
\bibfield{author}{\bibinfo{person}{Luca Comanducci}, \bibinfo{person}{Paolo
  Bestagini}, {and} \bibinfo{person}{Stefano Tubaro}.}
  \bibinfo{year}{2024}\natexlab{}.
\newblock \bibinfo{title}{{FakeMusicCaps}: A Dataset for Detection and
  Attribution of Synthetic Music Generated via Text-to-Music Models}.
\newblock
\showeprint[arxiv]{2409.10684}~[cs.SD]


\bibitem[Fan et~al\mbox{.}(2024a)]%
        {10506099}
\bibfield{author}{\bibinfo{person}{Cunhang Fan}, \bibinfo{person}{Mingming
  Ding}, \bibinfo{person}{Jianhua Tao}, \bibinfo{person}{Ruibo Fu},
  \bibinfo{person}{Jiangyan Yi}, \bibinfo{person}{Zhengqi Wen}, {and}
  \bibinfo{person}{Zhao Lv}.} \bibinfo{year}{2024}\natexlab{a}.
\newblock \showarticletitle{Dual-Branch Knowledge Distillation for Noise-Robust
  Synthetic Speech Detection}.
\newblock \bibinfo{journal}{\emph{IEEE/ACM Transactions on Audio, Speech, and
  Language Processing}}  \bibinfo{volume}{32} (\bibinfo{year}{2024}),
  \bibinfo{pages}{2453--2466}.
\newblock
\href{https://doi.org/10.1109/TASLP.2024.3389643}{doi:\nolinkurl{10.1109/TASLP.2024.3389643}}


\bibitem[Fan et~al\mbox{.}(2023)]%
        {FAN2023102988}
\bibfield{author}{\bibinfo{person}{Cunhang Fan}, \bibinfo{person}{Jun Xue},
  \bibinfo{person}{Shunbo Dong}, \bibinfo{person}{Mingming Ding},
  \bibinfo{person}{Jiangyan Yi}, \bibinfo{person}{Jinpeng Li}, {and}
  \bibinfo{person}{Zhao Lv}.} \bibinfo{year}{2023}\natexlab{}.
\newblock \showarticletitle{Subband Fusion of Complex Spectrogram for Fake
  Speech Detection}.
\newblock \bibinfo{journal}{\emph{Speech Communication}}  \bibinfo{volume}{155}
  (\bibinfo{year}{2023}), \bibinfo{pages}{102988}.
\newblock
\href{https://doi.org/10.1016/j.specom.2023.102988}{doi:\nolinkurl{10.1016/j.specom.2023.102988}}


\bibitem[Fan et~al\mbox{.}(2024b)]%
        {FAN2024106320}
\bibfield{author}{\bibinfo{person}{Cunhang Fan}, \bibinfo{person}{Jun Xue},
  \bibinfo{person}{Jianhua Tao}, \bibinfo{person}{Jiangyan Yi},
  \bibinfo{person}{Chenglong Wang}, \bibinfo{person}{Chengshi Zheng}, {and}
  \bibinfo{person}{Zhao Lv}.} \bibinfo{year}{2024}\natexlab{b}.
\newblock \showarticletitle{Spatial Reconstructed Local Attention {Res2Net}
  with {F0} Subband for Fake Speech Detection}.
\newblock \bibinfo{journal}{\emph{Neural Networks}}  \bibinfo{volume}{175}
  (\bibinfo{year}{2024}), \bibinfo{pages}{106320}.
\newblock
\href{https://doi.org/10.1016/j.neunet.2024.106320}{doi:\nolinkurl{10.1016/j.neunet.2024.106320}}


\bibitem[Huang et~al\mbox{.}(2026)]%
        {huang2026envtricascade}
\bibfield{author}{\bibinfo{person}{Hengyan Huang}, \bibinfo{person}{Xiaoxuan
  Guo}, \bibinfo{person}{Jiayi Zhou}, \bibinfo{person}{Yuankun Xie},
  \bibinfo{person}{Jian Liu}, \bibinfo{person}{Haonan Cheng},
  \bibinfo{person}{Long Ye}, {and} \bibinfo{person}{Qin Zhang}.}
  \bibinfo{year}{2026}\natexlab{}.
\newblock \bibinfo{title}{{EnvTriCascade}: An Environment-Aware Tri-Stage
  Cascaded Framework for {ESDD2} 2026 Challenge}.
\newblock
\showeprint[arxiv]{2605.18409}~[cs.SD]


\bibitem[India et~al\mbox{.}(2019)]%
        {india19_interspeech}
\bibfield{author}{\bibinfo{person}{Miquel India}, \bibinfo{person}{Pooyan
  Safari}, {and} \bibinfo{person}{Javier Hernando}.}
  \bibinfo{year}{2019}\natexlab{}.
\newblock \showarticletitle{Self Multi-Head Attention for Speaker Recognition}.
  In \bibinfo{booktitle}{\emph{Interspeech 2019}}. \bibinfo{pages}{4305--4309}.
\newblock
\href{https://doi.org/10.21437/Interspeech.2019-2616}{doi:\nolinkurl{10.21437/Interspeech.2019-2616}}


\bibitem[Kinnunen et~al\mbox{.}(2017)]%
        {kinnunen17_interspeech}
\bibfield{author}{\bibinfo{person}{Tomi Kinnunen}, \bibinfo{person}{Md.
  Sahidullah}, \bibinfo{person}{H{\'e}ctor Delgado},
  \bibinfo{person}{Massimiliano Todisco}, \bibinfo{person}{Nicholas Evans},
  \bibinfo{person}{Junichi Yamagishi}, {and} \bibinfo{person}{Kong~Aik Lee}.}
  \bibinfo{year}{2017}\natexlab{}.
\newblock \showarticletitle{The {ASVspoof 2017} Challenge: Assessing the Limits
  of Replay Spoofing Attack Detection}. In
  \bibinfo{booktitle}{\emph{Interspeech 2017}}. \bibinfo{pages}{2--6}.
\newblock
\href{https://doi.org/10.21437/Interspeech.2017-1111}{doi:\nolinkurl{10.21437/Interspeech.2017-1111}}


\bibitem[Loshchilov and Hutter(2017)]%
        {loshchilov2017decoupled}
\bibfield{author}{\bibinfo{person}{Ilya Loshchilov} {and}
  \bibinfo{person}{Frank Hutter}.} \bibinfo{year}{2017}\natexlab{}.
\newblock \showarticletitle{Decoupled Weight Decay Regularization}.
\newblock \bibinfo{journal}{\emph{arXiv preprint arXiv:1711.05101}}
  (\bibinfo{year}{2017}).
\newblock


\bibitem[Pasad et~al\mbox{.}(2021)]%
        {9688093}
\bibfield{author}{\bibinfo{person}{Ankita Pasad}, \bibinfo{person}{Ju-Chieh
  Chou}, {and} \bibinfo{person}{Karen Livescu}.}
  \bibinfo{year}{2021}\natexlab{}.
\newblock \showarticletitle{Layer-Wise Analysis of a Self-Supervised Speech
  Representation Model}. In \bibinfo{booktitle}{\emph{2021 IEEE Automatic
  Speech Recognition and Understanding Workshop}}. \bibinfo{pages}{914--921}.
\newblock
\href{https://doi.org/10.1109/ASRU51503.2021.9688093}{doi:\nolinkurl{10.1109/ASRU51503.2021.9688093}}


\bibitem[Pasad et~al\mbox{.}(2023)]%
        {10096149}
\bibfield{author}{\bibinfo{person}{Ankita Pasad}, \bibinfo{person}{Bowen Shi},
  {and} \bibinfo{person}{Karen Livescu}.} \bibinfo{year}{2023}\natexlab{}.
\newblock \showarticletitle{Comparative Layer-Wise Analysis of Self-Supervised
  Speech Models}. In \bibinfo{booktitle}{\emph{Proceedings of ICASSP}}.
  \bibinfo{pages}{1--5}.
\newblock
\href{https://doi.org/10.1109/ICASSP49357.2023.10096149}{doi:\nolinkurl{10.1109/ICASSP49357.2023.10096149}}


\bibitem[Shazeer(2020)]%
        {shazeer2020gluvariantsimprovetransformer}
\bibfield{author}{\bibinfo{person}{Noam Shazeer}.}
  \bibinfo{year}{2020}\natexlab{}.
\newblock \bibinfo{title}{{GLU} Variants Improve Transformer}.
\newblock
\showeprint[arxiv]{2002.05202}~[cs.LG]


\bibitem[Tak et~al\mbox{.}(2022a)]%
        {9746213}
\bibfield{author}{\bibinfo{person}{Hemlata Tak}, \bibinfo{person}{Madhu
  Kamble}, \bibinfo{person}{Jose Patino}, \bibinfo{person}{Massimiliano
  Todisco}, {and} \bibinfo{person}{Nicholas Evans}.}
  \bibinfo{year}{2022}\natexlab{a}.
\newblock \showarticletitle{{RawBoost}: A Raw Data Boosting and Augmentation
  Method Applied to Automatic Speaker Verification Anti-Spoofing}. In
  \bibinfo{booktitle}{\emph{Proceedings of ICASSP}}.
  \bibinfo{pages}{6382--6386}.
\newblock
\href{https://doi.org/10.1109/ICASSP43922.2022.9746213}{doi:\nolinkurl{10.1109/ICASSP43922.2022.9746213}}


\bibitem[Tak et~al\mbox{.}(2022b)]%
        {tak2022automaticspeakerverificationspoofing}
\bibfield{author}{\bibinfo{person}{Hemlata Tak}, \bibinfo{person}{Massimiliano
  Todisco}, \bibinfo{person}{Xin Wang}, \bibinfo{person}{Jee weon Jung},
  \bibinfo{person}{Junichi Yamagishi}, {and} \bibinfo{person}{Nicholas Evans}.}
  \bibinfo{year}{2022}\natexlab{b}.
\newblock \bibinfo{title}{Automatic Speaker Verification Spoofing and Deepfake
  Detection Using wav2vec 2.0 and Data Augmentation}.
\newblock
\showeprint[arxiv]{2202.12233}~[eess.AS]


\bibitem[Todisco et~al\mbox{.}(2019)]%
        {todisco19_interspeech}
\bibfield{author}{\bibinfo{person}{Massimiliano Todisco}, \bibinfo{person}{Xin
  Wang}, \bibinfo{person}{Ville Vestman}, \bibinfo{person}{Md. Sahidullah},
  \bibinfo{person}{H{\'e}ctor Delgado}, \bibinfo{person}{Andreas Nautsch},
  \bibinfo{person}{Junichi Yamagishi}, \bibinfo{person}{Nicholas Evans},
  \bibinfo{person}{Tomi~H. Kinnunen}, {and} \bibinfo{person}{Kong~Aik Lee}.}
  \bibinfo{year}{2019}\natexlab{}.
\newblock \showarticletitle{{ASVspoof 2019}: Future Horizons in Spoofed and
  Fake Audio Detection}. In \bibinfo{booktitle}{\emph{Interspeech 2019}}.
  \bibinfo{pages}{1008--1012}.
\newblock
\href{https://doi.org/10.21437/Interspeech.2019-2249}{doi:\nolinkurl{10.21437/Interspeech.2019-2249}}


\bibitem[Wang et~al\mbox{.}(2024)]%
        {wang24_asvspoof}
\bibfield{author}{\bibinfo{person}{Xin Wang}, \bibinfo{person}{H{\'e}ctor
  Delgado}, \bibinfo{person}{Hemlata Tak}, \bibinfo{person}{Jee weon Jung},
  \bibinfo{person}{Hye jin Shim}, \bibinfo{person}{Massimiliano Todisco},
  \bibinfo{person}{Ivan Kukanov}, \bibinfo{person}{Xuechen Liu},
  \bibinfo{person}{Md. Sahidullah}, \bibinfo{person}{Tomi~H. Kinnunen},
  \bibinfo{person}{Nicholas Evans}, \bibinfo{person}{Kong~Aik Lee}, {and}
  \bibinfo{person}{Junichi Yamagishi}.} \bibinfo{year}{2024}\natexlab{}.
\newblock \showarticletitle{{ASVspoof 5}: Crowdsourced Speech Data, Deepfakes,
  and Adversarial Attacks at Scale}. In \bibinfo{booktitle}{\emph{The Automatic
  Speaker Verification Spoofing Countermeasures Workshop}}.
  \bibinfo{pages}{1--8}.
\newblock
\href{https://doi.org/10.21437/ASVspoof.2024-1}{doi:\nolinkurl{10.21437/ASVspoof.2024-1}}


\bibitem[weon Jung et~al\mbox{.}(2022)]%
        {aasist}
\bibfield{author}{\bibinfo{person}{Jee weon Jung}, \bibinfo{person}{Hee-Soo
  Heo}, \bibinfo{person}{Hemlata Tak}, \bibinfo{person}{Hye jin Shim},
  \bibinfo{person}{Joon~Son Chung}, \bibinfo{person}{Bong-Jin Lee},
  \bibinfo{person}{Ha-Jin Yu}, {and} \bibinfo{person}{Nicholas Evans}.}
  \bibinfo{year}{2022}\natexlab{}.
\newblock \showarticletitle{{AASIST}: Audio Anti-Spoofing Using Integrated
  Spectro-Temporal Graph Attention Networks}. In
  \bibinfo{booktitle}{\emph{Proceedings of ICASSP}}. \bibinfo{publisher}{IEEE},
  \bibinfo{address}{Piscataway, NJ, USA}, \bibinfo{pages}{6367--6371}.
\newblock


\bibitem[Wu et~al\mbox{.}(2015)]%
        {wu15e_interspeech}
\bibfield{author}{\bibinfo{person}{Zhizheng Wu}, \bibinfo{person}{Tomi
  Kinnunen}, \bibinfo{person}{Nicholas Evans}, \bibinfo{person}{Junichi
  Yamagishi}, \bibinfo{person}{Cemal Hanil{\c{c}}i}, \bibinfo{person}{Md.
  Sahidullah}, {and} \bibinfo{person}{Aleksandr Sizov}.}
  \bibinfo{year}{2015}\natexlab{}.
\newblock \showarticletitle{{ASVspoof 2015}: The First Automatic Speaker
  Verification Spoofing and Countermeasures Challenge}. In
  \bibinfo{booktitle}{\emph{Interspeech 2015}}. \bibinfo{pages}{2037--2041}.
\newblock
\href{https://doi.org/10.21437/Interspeech.2015-462}{doi:\nolinkurl{10.21437/Interspeech.2015-462}}


\bibitem[Xie et~al\mbox{.}(2023a)]%
        {xie2023w2vasdg}
\bibfield{author}{\bibinfo{person}{Yuankun Xie}, \bibinfo{person}{Haonan
  Cheng}, \bibinfo{person}{Yutian Wang}, {and} \bibinfo{person}{Long Ye}.}
  \bibinfo{year}{2023}\natexlab{a}.
\newblock \showarticletitle{Learning A Self-Supervised Domain-Invariant Feature
  Representation for Generalized Audio Deepfake Detection}. In
  \bibinfo{booktitle}{\emph{Proceedings of Interspeech}}.
  \bibinfo{pages}{2808--2812}.
\newblock
\href{https://doi.org/10.21437/Interspeech.2023-1383}{doi:\nolinkurl{10.21437/Interspeech.2023-1383}}


\bibitem[Xie et~al\mbox{.}(2023b)]%
        {xie2023singledomain}
\bibfield{author}{\bibinfo{person}{Yuankun Xie}, \bibinfo{person}{Haonan
  Cheng}, \bibinfo{person}{Yutian Wang}, {and} \bibinfo{person}{Long Ye}.}
  \bibinfo{year}{2023}\natexlab{b}.
\newblock \showarticletitle{Single Domain Generalization for Audio Deepfake
  Detection}. In \bibinfo{booktitle}{\emph{Proceedings of the IJCAI 2023
  Workshop on Deepfake Audio Detection and Analysis}}
  \emph{(\bibinfo{series}{CEUR Workshop Proceedings},
  Vol.~\bibinfo{volume}{3597})}. \bibinfo{publisher}{CEUR-WS.org},
  \bibinfo{pages}{58--63}.
\newblock


\bibitem[Xie et~al\mbox{.}(2024a)]%
        {xie2024asdg}
\bibfield{author}{\bibinfo{person}{Yuankun Xie}, \bibinfo{person}{Haonan
  Cheng}, \bibinfo{person}{Yutian Wang}, {and} \bibinfo{person}{Long Ye}.}
  \bibinfo{year}{2024}\natexlab{a}.
\newblock \showarticletitle{Domain Generalization via Aggregation and
  Separation for Audio Deepfake Detection}.
\newblock \bibinfo{journal}{\emph{IEEE Transactions on Information Forensics
  and Security}}  \bibinfo{volume}{19} (\bibinfo{year}{2024}),
  \bibinfo{pages}{344--358}.
\newblock
\href{https://doi.org/10.1109/TIFS.2023.3324724}{doi:\nolinkurl{10.1109/TIFS.2023.3324724}}


\bibitem[Xie et~al\mbox{.}(2026a)]%
        {atadd2026}
\bibfield{author}{\bibinfo{person}{Yuankun Xie}, \bibinfo{person}{Haonan
  Cheng}, \bibinfo{person}{Jiayi Zhou}, \bibinfo{person}{Xiaoxuan Guo},
  \bibinfo{person}{Tao Wang}, \bibinfo{person}{Jian Liu},
  \bibinfo{person}{Weiqiang Wang}, \bibinfo{person}{Ruibo Fu},
  \bibinfo{person}{Xiaopeng Wang}, \bibinfo{person}{Hengyan Huang},
  \bibinfo{person}{Xiaoying Huang}, \bibinfo{person}{Long Ye}, {and}
  \bibinfo{person}{Guangtao Zhai}.} \bibinfo{year}{2026}\natexlab{a}.
\newblock \showarticletitle{{AT-ADD}: All-Type Audio Deepfake Detection
  Challenge Evaluation Plan}. In \bibinfo{booktitle}{\emph{Proceedings of the
  ACM International Conference on Multimedia}}.
\newblock
\shownote{ACM Multimedia 2026 Grand Challenge}.
\newblock


\bibitem[Xie et~al\mbox{.}(2026b)]%
        {xie2026wpt}
\bibfield{author}{\bibinfo{person}{Yuankun Xie}, \bibinfo{person}{Ruibo Fu},
  \bibinfo{person}{Xiaopeng Wang}, \bibinfo{person}{Zhiyong Wang},
  \bibinfo{person}{Songjun Cao}, \bibinfo{person}{Long Ma},
  \bibinfo{person}{Haonan Cheng}, {and} \bibinfo{person}{Long Ye}.}
  \bibinfo{year}{2026}\natexlab{b}.
\newblock \showarticletitle{Detect All-Type Deepfake Audio: Wavelet Prompt
  Tuning for Enhanced Auditory Perception}.
\newblock \bibinfo{journal}{\emph{Proceedings of the AAAI Conference on
  Artificial Intelligence}} \bibinfo{volume}{40}, \bibinfo{number}{42}
  (\bibinfo{year}{2026}), \bibinfo{pages}{35922--35930}.
\newblock
\href{https://doi.org/10.1609/aaai.v40i42.40907}{doi:\nolinkurl{10.1609/aaai.v40i42.40907}}


\bibitem[Xie et~al\mbox{.}(2026c)]%
        {xie2026fsw}
\bibfield{author}{\bibinfo{person}{Yuankun Xie}, \bibinfo{person}{Ruibo Fu},
  \bibinfo{person}{Xiaopeng Wang}, \bibinfo{person}{Zhiyong Wang},
  \bibinfo{person}{Ya Li}, \bibinfo{person}{Yingming Gao},
  \bibinfo{person}{Zhengqi Wen}, \bibinfo{person}{Haonan Cheng}, {and}
  \bibinfo{person}{Long Ye}.} \bibinfo{year}{2026}\natexlab{c}.
\newblock \showarticletitle{Fake Speech Wild: Detecting Deepfake Speech on
  Social Media Platform}. In \bibinfo{booktitle}{\emph{Proceedings of the IEEE
  International Conference on Acoustics, Speech and Signal Processing}}.
\newblock
\href{https://doi.org/10.1109/ICASSP55912.2026.11461877}{doi:\nolinkurl{10.1109/ICASSP55912.2026.11461877}}


\bibitem[Xie et~al\mbox{.}(2024b)]%
        {xie2024refd}
\bibfield{author}{\bibinfo{person}{Yuankun Xie}, \bibinfo{person}{Ruibo Fu},
  \bibinfo{person}{Zhengqi Wen}, \bibinfo{person}{Zhiyong Wang},
  \bibinfo{person}{Xiaopeng Wang}, \bibinfo{person}{Haonnan Cheng},
  \bibinfo{person}{Long Ye}, {and} \bibinfo{person}{Jianhua Tao}.}
  \bibinfo{year}{2024}\natexlab{b}.
\newblock \showarticletitle{Generalized Source Tracing: Detecting Novel Audio
  Deepfake Algorithm with Real Emphasis and Fake Dispersion Strategy}. In
  \bibinfo{booktitle}{\emph{Proceedings of Interspeech}}.
  \bibinfo{pages}{4833--4837}.
\newblock
\href{https://doi.org/10.21437/Interspeech.2024-254}{doi:\nolinkurl{10.21437/Interspeech.2024-254}}


\bibitem[Xie et~al\mbox{.}(2024c)]%
        {xie2024fsd}
\bibfield{author}{\bibinfo{person}{Yuankun Xie}, \bibinfo{person}{Jingjing
  Zhou}, \bibinfo{person}{Xiaolin Lu}, \bibinfo{person}{Zhenghao Jiang},
  \bibinfo{person}{Yuxin Yang}, \bibinfo{person}{Haonan Cheng}, {and}
  \bibinfo{person}{Long Ye}.} \bibinfo{year}{2024}\natexlab{c}.
\newblock \showarticletitle{{FSD}: An Initial Chinese Dataset for Fake Song
  Detection}. In \bibinfo{booktitle}{\emph{Proceedings of the IEEE
  International Conference on Acoustics, Speech and Signal Processing}}.
  \bibinfo{pages}{4605--4609}.
\newblock
\href{https://doi.org/10.1109/ICASSP48485.2024.10446271}{doi:\nolinkurl{10.1109/ICASSP48485.2024.10446271}}


\bibitem[Xue et~al\mbox{.}(2026a)]%
        {xue2026unifyingspeecheditingdetection}
\bibfield{author}{\bibinfo{person}{Jun Xue}, \bibinfo{person}{Yi Chai},
  \bibinfo{person}{Yanzhen Ren}, \bibinfo{person}{Jinshen He},
  \bibinfo{person}{Zhiqiang Tang}, \bibinfo{person}{Zhuolin Yi},
  \bibinfo{person}{Yihuan Huang}, \bibinfo{person}{Yuankun Xie}, {and}
  \bibinfo{person}{Yujie Chen}.} \bibinfo{year}{2026}\natexlab{a}.
\newblock \bibinfo{title}{Unifying Speech Editing Detection and Content
  Localization via Prior-Enhanced Audio {LLM}s}.
\newblock
\showeprint[arxiv]{2601.21463}~[cs.SD]


\bibitem[Xue et~al\mbox{.}(2023)]%
        {10096837}
\bibfield{author}{\bibinfo{person}{Jun Xue}, \bibinfo{person}{Cunhang Fan},
  \bibinfo{person}{Jiangyan Yi}, \bibinfo{person}{Chenglong Wang},
  \bibinfo{person}{Zhengqi Wen}, \bibinfo{person}{Dan Zhang}, {and}
  \bibinfo{person}{Zhao Lv}.} \bibinfo{year}{2023}\natexlab{}.
\newblock \showarticletitle{Learning From Yourself: A Self-Distillation Method
  for Fake Speech Detection}. In \bibinfo{booktitle}{\emph{Proceedings of
  ICASSP}}. \bibinfo{pages}{1--5}.
\newblock
\href{https://doi.org/10.1109/ICASSP49357.2023.10096837}{doi:\nolinkurl{10.1109/ICASSP49357.2023.10096837}}


\bibitem[Xue et~al\mbox{.}(2024)]%
        {10605999}
\bibfield{author}{\bibinfo{person}{Jun Xue}, \bibinfo{person}{Cunhang Fan},
  \bibinfo{person}{Jiangyan Yi}, \bibinfo{person}{Jian Zhou}, {and}
  \bibinfo{person}{Zhao Lv}.} \bibinfo{year}{2024}\natexlab{}.
\newblock \showarticletitle{Dynamic Ensemble Teacher-Student Distillation
  Framework for Light-Weight Fake Audio Detection}.
\newblock \bibinfo{journal}{\emph{IEEE Signal Processing Letters}}
  \bibinfo{volume}{31} (\bibinfo{year}{2024}), \bibinfo{pages}{2305--2309}.
\newblock
\href{https://doi.org/10.1109/LSP.2024.3431936}{doi:\nolinkurl{10.1109/LSP.2024.3431936}}


\bibitem[Xue et~al\mbox{.}(2026b)]%
        {xue2026rtcfakespeechdeepfakedetection}
\bibfield{author}{\bibinfo{person}{Jun Xue}, \bibinfo{person}{Zhuolin Yi},
  \bibinfo{person}{Yihuan Huang}, \bibinfo{person}{Yanzhen Ren},
  \bibinfo{person}{Yujie Chen}, \bibinfo{person}{Cunhang Fan},
  \bibinfo{person}{Zicheng Su}, \bibinfo{person}{Yonghong Zhang}, {and}
  \bibinfo{person}{Bo Cai}.} \bibinfo{year}{2026}\natexlab{b}.
\newblock \bibinfo{title}{{RTCFake}: Speech Deepfake Detection in Real-Time
  Communication}.
\newblock
\showeprint[arxiv]{2604.23742}~[cs.SD]


\bibitem[Yamagishi et~al\mbox{.}(2021)]%
        {yamagishi21_asvspoof}
\bibfield{author}{\bibinfo{person}{Junichi Yamagishi}, \bibinfo{person}{Xin
  Wang}, \bibinfo{person}{Massimiliano Todisco}, \bibinfo{person}{Md.
  Sahidullah}, \bibinfo{person}{Jose Patino}, \bibinfo{person}{Andreas
  Nautsch}, \bibinfo{person}{Xuechen Liu}, \bibinfo{person}{Kong~Aik Lee},
  \bibinfo{person}{Tomi Kinnunen}, \bibinfo{person}{Nicholas Evans}, {and}
  \bibinfo{person}{H{\'e}ctor Delgado}.} \bibinfo{year}{2021}\natexlab{}.
\newblock \showarticletitle{{ASVspoof 2021}: Accelerating Progress in Spoofed
  and Deepfake Speech Detection}. In \bibinfo{booktitle}{\emph{2021 Edition of
  the Automatic Speaker Verification and Spoofing Countermeasures Challenge}}.
  \bibinfo{pages}{47--54}.
\newblock
\href{https://doi.org/10.21437/ASVSPOOF.2021-8}{doi:\nolinkurl{10.21437/ASVSPOOF.2021-8}}


\bibitem[Yi et~al\mbox{.}(2022)]%
        {add2022}
\bibfield{author}{\bibinfo{person}{Jiangyan Yi}, \bibinfo{person}{Ruibo Fu},
  \bibinfo{person}{Jianhua Tao}, \bibinfo{person}{Shuai Nie},
  \bibinfo{person}{Haoxin Ma}, \bibinfo{person}{Chenglong Wang},
  \bibinfo{person}{Tao Wang}, \bibinfo{person}{Zhengkun Tian},
  \bibinfo{person}{Ye Bai}, \bibinfo{person}{Cunhang Fan},
  \bibinfo{person}{Shan Liang}, \bibinfo{person}{Shiming Wang},
  \bibinfo{person}{Shuai Zhang}, \bibinfo{person}{Xinrui Yan},
  \bibinfo{person}{Le Xu}, \bibinfo{person}{Zhengqi Wen},
  \bibinfo{person}{Haizhou Li}, \bibinfo{person}{Zheng Lian}, {and}
  \bibinfo{person}{Bin Liu}.} \bibinfo{year}{2022}\natexlab{}.
\newblock \showarticletitle{{ADD 2022}: The First Audio Deep Synthesis
  Detection Challenge}. In \bibinfo{booktitle}{\emph{Proceedings of ICASSP}}.
  \bibinfo{publisher}{IEEE}, \bibinfo{address}{Piscataway, NJ, USA},
  \bibinfo{pages}{9216--9220}.
\newblock


\bibitem[Yi et~al\mbox{.}(2023)]%
        {yi2023add2023secondaudio}
\bibfield{author}{\bibinfo{person}{Jiangyan Yi}, \bibinfo{person}{Jianhua Tao},
  \bibinfo{person}{Ruibo Fu}, \bibinfo{person}{Xinrui Yan},
  \bibinfo{person}{Chenglong Wang}, \bibinfo{person}{Tao Wang},
  \bibinfo{person}{Chu~Yuan Zhang}, \bibinfo{person}{Xiaohui Zhang},
  \bibinfo{person}{Yan Zhao}, \bibinfo{person}{Yong Ren}, \bibinfo{person}{Le
  Xu}, \bibinfo{person}{Junzuo Zhou}, \bibinfo{person}{Hao Gu},
  \bibinfo{person}{Zhengqi Wen}, \bibinfo{person}{Shan Liang},
  \bibinfo{person}{Zheng Lian}, \bibinfo{person}{Shuai Nie}, {and}
  \bibinfo{person}{Haizhou Li}.} \bibinfo{year}{2023}\natexlab{}.
\newblock \bibinfo{title}{{ADD 2023}: The Second Audio Deepfake Detection
  Challenge}.
\newblock
\showeprint[arxiv]{2305.13774}~[cs.SD]


\bibitem[Yin et~al\mbox{.}(2025)]%
        {yin25_interspeech}
\bibfield{author}{\bibinfo{person}{Han Yin}, \bibinfo{person}{Yang Xiao},
  \bibinfo{person}{Rohan~Kumar Das}, \bibinfo{person}{Jisheng Bai},
  \bibinfo{person}{Haohe Liu}, \bibinfo{person}{Wenwu Wang}, {and}
  \bibinfo{person}{Mark~D. Plumbley}.} \bibinfo{year}{2025}\natexlab{}.
\newblock \showarticletitle{{EnvSDD}: Benchmarking Environmental Sound Deepfake
  Detection}. In \bibinfo{booktitle}{\emph{Interspeech 2025}}.
  \bibinfo{pages}{201--205}.
\newblock
\href{https://doi.org/10.21437/Interspeech.2025-1143}{doi:\nolinkurl{10.21437/Interspeech.2025-1143}}


\bibitem[Zhang et~al\mbox{.}(2024a)]%
        {xwsb}
\bibfield{author}{\bibinfo{person}{Qishan Zhang}, \bibinfo{person}{Shuangbing
  Wen}, \bibinfo{person}{Fangke Yan}, \bibinfo{person}{Tao Hu}, {and}
  \bibinfo{person}{Jun Li}.} \bibinfo{year}{2024}\natexlab{a}.
\newblock \showarticletitle{{XWSB}: A Blend System Utilizing XLS-R and WavLM
  With SLS Classifier Detection System for SVDD 2024 Challenge}. In
  \bibinfo{booktitle}{\emph{2024 IEEE Spoken Language Technology Workshop}}.
  \bibinfo{publisher}{IEEE}, \bibinfo{address}{Piscataway, NJ, USA},
  \bibinfo{pages}{788--794}.
\newblock


\bibitem[Zhang et~al\mbox{.}(2024b)]%
        {svdd}
\bibfield{author}{\bibinfo{person}{You Zhang}, \bibinfo{person}{Yongyi Zang},
  \bibinfo{person}{Jiatong Shi}, \bibinfo{person}{Ryuichi Yamamoto},
  \bibinfo{person}{Tomoki Toda}, {and} \bibinfo{person}{Zhiyao Duan}.}
  \bibinfo{year}{2024}\natexlab{b}.
\newblock \showarticletitle{{SVDD 2024}: The Inaugural Singing Voice Deepfake
  Detection Challenge}. In \bibinfo{booktitle}{\emph{2024 IEEE Spoken Language
  Technology Workshop}}. \bibinfo{publisher}{IEEE},
  \bibinfo{address}{Piscataway, NJ, USA}, \bibinfo{pages}{782--787}.
\newblock


\end{thebibliography}
